\documentstyle[12pt]{article}
\def\b{\begin{equation}}
\def\e{\end{equation}}
\def\ba{\begin{eqnarray}}
\def\ea{\end{eqnarray}}
\def\sla{\hbox{$\!\!\!/$}}

\def\b{\begin{equation}}
\def\e{\end{equation}}
\def\ba{\begin{eqnarray}}
\def\ea{\end{eqnarray}}
\def\sla{\hbox{$\!\!\!/$}}

\begin{document}

\title{Consistency in Regularizations of the Gauged NJL Model at the One Loop Level
$^{*}$}
\author{O.A. Battistel$^{(1)**}$ and M.C. Nemes$^{(1)}$\\ 
(1) Departamento de F\'{\i}sica,\\ 
 Univ. Federal de Minas Gerais, UFMG\\
 CP 702, CEP 30161-970 Belo Horizonte, MG, Brasil}
\date{\today}
\maketitle

\centerline{{\bf ABSTRACT}}
In this work we revisit questions recently raised in the literature 
associated to relevant but divergent amplitudes in the 
gauged NJL model. The questions raised involve ambiguities and symmetry 
violations which concern the model's predictive power at one loop level.
Our study shows by means of an alternative prescription 
to handle divergent amplitudes, that it is possible to obtain unambiguous and 
symmetry preserving amplitudes. The procedure adopted makes use solely of 
{\it general} properties of an eventual regulator, thus avoiding an explicit 
form. We find, after a thorough analysis of the problem that there are well 
established conditions to be fulfiled by any consistent regularization
prescription in order to avoid the problems of concern at one loop level.


[Keywords: Ward Identities for two and three point functions,
Nambu-Jona-Lasinio Model]

\footnotesize{
*This work was partially supported by CNPq.

**On leave from Univ. Federal de Santa Maria, Santa Maria, Rs, Brasil, 
CP 5093, CEP 97119-900, 
e-mail: orimar@super.ufsm.br}

\normalsize

\newpage

\section{Introduction}
The Nambu-Jona-Lasinio model ${\cite{ref1}}$ has been extensively
studied in the context of low energy hadron physics
${\cite{ref2}}$, ${\cite{ref3}}$, ${\cite{ref4}}$, ${\cite{ref5}}$.
Due the presence of the four fermion interaction the model is
nonrenormalizable in he weak coupling expansion for d>2 dimensions.
Therefore a regularization scheme is called for whenever divergent
amplitudes appear. Frequently, regularization prescriptions adopted in
the calculation of divergent amplitudes destroy the symmetries of the
original lagrangean and introduce nonphysical behavior like unitarity
violation or unphysical thresholds. This is unavoidable if finite integrals
are regularized. An elegant way to circumvent this problem has been
proposed in ${\cite{ref6}}$ where dispersion relations are used.
This however requires the extension of all integrals to infinity. This can
be justified at the one loop level, since, as has been shown in refs.
${\cite{ref7}}$, ${\cite{ref8}}$, ${\cite{ref9}}$, the NJL constitutes a
perfectly renormalizable field theory in the mean of field expansion,
also for d>2. This naturally avoids unitarity violation and the appearance
of nonphysical thresholds. There remains, however, the problem of
ambiguities  and symmetry violations. In fact, Ward identities have
been used to constrain regularization of the NJL model ${\cite{ref10}}$,
${\cite{ref11}}$. In the present work we want to investigate the source of
ambiguities and symmetry violations in the NJL in a way which is as little
as possible committed to a given regularization scheme. We therefore only
assume the existence of a implicit regularization scheme and derive the
properties it should have in order to avoid such problems. The existence
of such scheme has been proven in ref ${\cite{ref12}}$ and given here in
Appendix A, for the sake of completeness.

The context of our discussion and physical motivation is the following :
Recently the NJL model has been used to provide for a possible mechanism
to generate dynamical
symmetry breaking in the context of the Standard Model. The reason for
this being that the top quark mass is much heavier than that of the gauge
bosons. Bardeen, Hill and Lindner (BHL) used the NJL model, in one loop
approximation, in order to predict the top quark and the Higgs Boson
masses ${\cite{ref13}}$. Given the nonrenormalizability of the NJL
model, the calculations made use of a particular regularization scheme.
Latter, Wiley ${\cite{ref14}}$ argued that the regularization scheme
used by BHL turns the evaluated amplitudes ambiguous, due to their
divergent character. The conclusion of Wiley's work is that the BHL
results are consequently not consistent.

More recently T. Ghergheta ${\cite{ref6}}$ returned to this discussion.
emphasizing that the qualitative equivalence of the NJL model and the
Standard Model is strongly dependent on the choice of an adequate
regularization scheme. The author argues that the sharp cutoff method is
not consistent since it breaks gauge invariance and introduces
ambiguities related to the choices of momentum routing in the internal
lines of the loops. On the other hand Dimensional Regularization,
inspite of being a consistent prescription from the point of
view of symmetry considerations and ambiguities,
cannot be used since it eliminates the quadratic divergence, essential for
the model. The problems were
circumvented in ref. ${\cite{ref6}}$ by using dispersion relations and
Cutkosky's rules.

In the present paper we revise the question of the predictive power of
the NJL model at one loop level. Having this in mind we adopt a different
strategy for the 
manipulation and calculation of divergent amplitudes which clearly displays 
the sources of all ambiguities and symmetry violations in a way which is 
independent of the specific regularization prescription. We show that there 
exists very general conditions to be obeyed by {\it any} regularization 
prescription in order to obtain consistent results.

Section II contains the explicit calculation of by 
point functions necessary for the discussion of ref. ${\cite{ref6}}$.
We test the prescription verifying all Ward Identities
related to the amplitudes and by considering the possibility of
ambiguities. In the section III we present the  conclusions and final remarks.

\section{Two Point Functions and Ambiguities}
The first two point function to be considered is the scalar-scalar one,
which is necessary for the description of the {\it t $\bar{t}$} channel of the
fermion-fermion scattering amplitude. It is defined by
\b
T^{SS} =\int \frac{d^4k}{(2\pi )^4}Tr \left\{ \hat{1} \frac{1}
{[(k\sla +k\sla_1) -m_t]}\hat{1} \frac{1}{[(k\sla +k\sla _2)
-m_t]}\right\}
\e
where $m_t$ is the top quark mass, $k_1$ and $k_2$ are arbitrary
internal momentum routings. The choice $k_1=(1-\alpha )p$ and
$k_2=-\alpha p$ corresponds to that of ${\cite{ref6}}$. Note that only
the difference $k_1-k_2$ is a physical quantity (external momentum $q$).
The sum $k_1+k_2$ or the product $k_1k_2$ are ambiguous quantities.

After taking the Dirac trace and reorganizing the expression we get 
\begin{eqnarray}
T^{SS}&=&2\left\{\int\frac{d^4k}{(2\pi )^4}\frac{1}
{[(k+k_1)^2-m_t^2]}\right. \nonumber \\
& &+\int\frac{d^4k}{(2\pi )^4}\frac{1}
{[(k+k_2)^2-m_t^2]} \nonumber \\
& &+\left.[4m_t^2-(k_1-k_2)^2]\int\frac{d^4k}{(2\pi )^4}\frac{1}
{[(k+k_1)^2-m_t^2][(k+k_2)^2-m_t^2]} \right\},
\end{eqnarray}

At this point the usual procedure is to adopt a regularization scheme.
Instead of doing this at this stage we adopt a different strategy. 
We implicitly assume some generic regulating function in
all steps and indicate with the letter $\Lambda$ under the integral sign. The 
existence of the connection limit is used for removing the subscript (regularization) 
from finite integrals. 
All we need from this function is that it is an even function
of loop momentum and that a connection limit exists.

We first consider the quadratically divergent integral, which we
reorganize using a convenient identity at the level of the integrands

\begin{eqnarray}
\int_\Lambda \frac{d^4k}{(2\pi )^4}\frac{1 }{[(k +k_1)^2 -m_t^2 ]}&=&
\int_\Lambda \frac{d^4k}{(2\pi )^4}\frac{1 }{(k^2 -m_t^2 )} \nonumber \\
& &+k_{1\alpha} k_{1\beta} \left\{\int_\Lambda \frac{d^4k}{(2\pi )^4}
\frac{4k_\alpha k_\beta }{(k^2 -m_t^2 )^3}
-\int_\Lambda \frac{d^4k}{(2\pi )^4}\frac{\delta_{\alpha\beta}}{(k^2 -m_t^2 )^2} 
\right\}\nonumber \\
& &\!\!\!\!\!\!\!\!\!\!\!\!\!\!\!\!\!\!\!\!\!\!\!\!+\left\{\int_\Lambda \frac{d^4k}{(2\pi )^4}\frac{(k_1)^2}{(k^2 -m_t^2 )^3} 
-\int_\Lambda 
\frac{d^4k}{(2\pi )^4}\frac{( k_1^2+2k_1\cdot k)^3
}{(k^2 -m_t^2 )^3[(k+k_1)^2-m_t^2]} \right\}.
\end{eqnarray}

The last two integrals thus obtained are finite. They are integrated
without restrictions. It is precisely at this point that nonphysical
thresholds and unitarity violations are introduced by usual
regularization procedures, i.e., by modifying the external momentum
dependence of the finite (physical!) part of the amplitudes. The
direct integration, in this case, yields an important exact cancellation.
The remaining integrals will be left as they appear. In this specific
case our philosophy is equivalent in spirit to the usual BPHZ procedure
which makes use of successive subtractions around a
fixed external momentum. It is worth noticing that the arbitrary
choice for the internal lines does not allows us to attribute any
physical meaning to $k_1$ and $k_2$. As will become clear in what
follows the difference between our procedure and the BPHZ subtraction
scheme in the case of different masses are still more marked. From our
point of view any convenient identity can be used. Taylor expansions
are a possible choice when adequate.

The other integral logarithimically divergent, may be reorganized as
follows 
\begin{eqnarray}
\int_\Lambda \frac{d^4k}{(2\pi )^4}\frac{1}{[(k+k_1)^2-m_t^2]
[(k+k_2)^2-m_t^2]}&=&
\int_\Lambda \frac{d^4k}{(2\pi )^4}\frac{1}
{(k^2-m_t^2)^2}\nonumber \\
& &-\int\frac{d^4k}{(2\pi )^4}\frac{(k_1^2+2k_1\cdot k)}
{(k^2-m_t^2)^2[(k+k_1)^2-m_t^2]}\nonumber \\
& &-\int\frac{d^4k}{(2\pi )^4}\frac{(k_2^2+2k_2\cdot k)}
{(k^2-m_t^2)^2[(k+k_2)^2-m_t^2]}\nonumber \\
& &\!\!\!\!\!\!\!\!\!\!\!\!\!\!\!\!\!\!\!\!\!\!\!\!\!\!\!\!\!\!\!\!\!\!\!\!\!\!\!\!\!\!\!\!\!\!\!\!\!\!\!\!\!
+\int\frac{d^4k}{(2\pi )^4}\frac{(k_1^2+2k_1\cdot k)(k_2^2+2k_2\cdot k)}
{(k^2-m_t^2)^2[(k+k_1)^2-m_t^2][(k+k_2)^2-m_t^2]}
\label{eq24}
\end{eqnarray}
This identity is not unique but is convenient to maintain the symmetry
between $k_1$ and $k_2$ explicitly. Now we
perform the three last
integrals to obtain
\begin{equation}
\int_\Lambda \frac{d^4k}{(2\pi )^4}\frac{1}
{[(k+k_1)^2-m_t^2][(k+k_2)^2-m_t^2]}=I_{log}(m_t^2)-
\left(\frac{i}{(4\pi )^2}\right)\left[
Z_0(m_t^2,m_t^2;(k_1-k_2)^2;m_t^2)\right]
\label{eq25}
\end{equation}
where we have introduced the definitions, the basic divergent
logarithmic object
\b
I_{log}(m_t^2)=\int_\Lambda \frac{d^4k}{(2\pi )^4}\frac{1}{(k^2 -m_t^2)^2},
\e
and the one loop structure function ${\cite{ref15}}$
\begin{equation}
Z_k(\lambda_1^2,\lambda_2^2;q^2;\lambda^2)=\int_0^1dz z^k \ln{\left(
\frac{q^2z(1-z)+(\lambda_1^2-\lambda_2^2)z
-\lambda^2_1}{-\lambda^2}\right)}.
\e

Collecting all results together we have, taking $(k_1-k_2)\equiv q$,
\ba
T^{SS}&=&4\left\{I_{quad}(m_t^2)+\frac{[4m_t^2-q^2]}{2}I_{log}(m_t^2)
+\frac{[4m_t^2-q^2]}{2}
Z_k(m_t^2,m_t^2;q^2;m_t^2)\right\}\nonumber \\
& &+2k_{1\alpha}k_{1\beta}\triangle_{\alpha\beta}(m_t^2)
+2k_{2\alpha}k_{2\beta}\triangle_{\alpha\beta}(m_t^2)
\ea
where we define the basic quadratically divergent object
\b
I_{quad}(m_t^2)=\int_\Lambda \frac{d^4k}{(2\pi )^4}\frac{1}{(k^2 -m_t^2)},
\e
and
\b
\triangle_{\alpha\beta}=\int_\Lambda \frac{d^4k}{(2\pi )^4}\frac{4k_\alpha 
k_\beta }{(k^2 -m_t^2
)^3}-\int_\Lambda \frac{d^4k}{(2\pi )^4}\frac{g_{\alpha\beta}}{(k^2 -m_t^2
)^2}.
\e

We stress that the results of ref${\cite{ref6}}$, eq.(6), is still
contained in the above results. We will return to this point later.

The next two point function to be considered is the Pseudoscalar-Pseudoscalar, 
defined by 
\begin{equation}
T^{PP} =\int \frac{d^4k}{(2\pi )^4}Tr \left\{\gamma_5 \frac{1}
{[k\sla +k\sla _1-m_t]}\gamma_5 \frac{1}{[k\sla +k\sla _2 -m_t]}\right\},
\end{equation}
which is necessary in the neutral $t\bar{t}$ channel of the scattering
amplitude. Using the same ingredients as in previous calculation we obtain:
\ba
T^{PP} &= &
4\left\{(-)[I_{quad}(m_t^2)]+\frac{q^2}{2}[I_{log}(m_t^2)]\right.
\nonumber \\
& &\left.-\left(\frac{i}{(4\pi)^2}\right)\frac{q^2}{2}
[Z_0(m_t^2,m_t^2,q^2;m_t^2]
\right\}\nonumber \\
& &-2k_{1\alpha}k_{1\beta}[\triangle_{\alpha\beta}(m_t^2)]
-2k_{2\alpha}k_{2\beta}[\triangle_{\alpha\beta}(m_t^2)].
\ea

At this point most of the calculations with the NJL including that of ref${\cite{ref6}}$
use the gap equation to replace the quadratic divergence. However if the matter is 
ambiguities it is important to question, how unambiguous is the gap equation 
itself? In principle nothing prevents us from using an arbitrary momentum routing 
also in the scalar one point function, which originates the gap equation. Defining the 
one point scalar function as:
\begin{equation}
T^S(m_t^2) =\int \frac{d^4k}{(2\pi )^4}Tr\left\{\hat{1}\frac{1}
{[k\sla +l\sla -m_t ]}\right\},
\end{equation}
we get for it
\begin{equation}
T^S(m_t^2) =4m_t\int \frac{d^4k}{(2\pi )^4}\frac{1}
{(k+l)^2 -m_t^2},
\end{equation}
where $l$ is an arbitrary internal momentum. The notation adopted indicates 
the mass carried by the propagator of the internal line. As we can see, this leads 
to an ambiguity in the gap equation itself if the result of this integral is 
$l$-dependent. Using the results obtained previously in this integral:
\b
T^S=4m_t\left\{I_{quad}(m_t^2)+l_{\alpha}l_{\beta}\triangle_{\alpha\beta}
(m_t^2)\right\}
\e
Up to this point the use of an gap equation in the two point function may be
mathematically dangerous from the point of view of ambiguities.

So far we have not yet made use of any regularization scheme. We have simply noticed 
a certain regularity in the form of the ambiguous terms. One could at this 
point search for a regularization scheme capable of eliminating all these 
ambiguities. However, our argument is that ambiguities are not the only 
problematic point to be circumvented in order to have the full predictive
power of the underlying model still present in the so far calculated 
quadratically divergent amplitudes. The other only major aspect to be considered 
is the {\it symmetry relations} which involve all necessary two point functions. 
Therefore we next consider all two point functions which are necessary to construct the gauge 
vector boson propagator for (eq.(9) of ref[4])
\b
\frac{1}{g^2_2}\left[D^\omega_{\mu\nu}(q)\right]^{-1}=\frac{1}{g^2_2}
(q_\mu q_\nu -g_{\mu\nu}q^2)+\frac{\Pi _{\mu\nu}(q)}{8}-\frac{1}{8}J_\mu (q) 
\Gamma_F(q^2)J_\nu (q)
\e
where $g_2$ is the $SU(2)$ coupling constant and
\b
\Gamma_F(q^2) =i\int \frac{d^4k}{(2\pi )^4}Tr \left\{ (1-\gamma_5) \frac{1}
{[(k\sla +k\sla_1) -m_t]}(1+\gamma_5) \frac{1}{[(k\sla +k\sla _2)
-m_b]}\right\}
\e

\b
J_\mu (q) =i\int \frac{d^4k}{(2\pi )^4}Tr \left\{ \gamma_\mu (1-\gamma_5) \frac{1}
{[(k\sla +k\sla_1) -m_t]}(1+\gamma_5) \frac{1}{[(k\sla +k\sla _2)
-m_b]}\right\}
\e

\b
J_{\mu\nu} (q) =i\int \frac{d^4k}{(2\pi )^4}Tr \left\{ \gamma_\mu (1-\gamma_5) \frac{1}
{[(k\sla +k\sla_1) -m_t]}\gamma_\nu (1+\gamma_5) \frac{1}{[(k\sla +k\sla _2)
-m_b]}\right\}
\e

Thus we need to calculate $T^{SS}$, $T^{PP}$, $T^{SP}$, $T_\mu^{VS}$, 
$T_\mu^{VP}$, $T_\mu^{AS}$, $T_\mu^{AP}$, $T_{\mu\nu}^{VV}$ and $T_{\mu\nu}^{AA}$. 
Let us then calculate these ingredients. Taking $T^{PP}$ (eq.(11)), with 
different masses now, after Dirac trace and some reorganization we get

\begin{eqnarray}
T^{PP}&=&2\left\{(-)\int\frac{d^4k}{(2\pi )^4}\frac{1}
{[(k+k_1)^2-m_t^2]}\right. \nonumber \\
& &-\int\frac{d^4k}{(2\pi )^4}\frac{1}
{[(k+k_2)^2-m_b^2]} \nonumber \\
& &+\left.[(m_t-m_b)^2-(k_1-k_2)^2]\int\frac{d^4k}{(2\pi )^4}\frac{1}
{[(k+k_1)^2-m_t^2][(k+k_2)^2-m_b^2]} \right\},
\end{eqnarray}

Note that the propagator that carries mass $m_t$ is rotulated by $k_1$ and those 
with mass $m_b$ rotulated by $k_2$.

The important question now is: how to perform the separation of the divergent 
and finite parts? Specifically, which one of the masses will be left in the divergent 
objects? This question is related to an important matter related to the manipulation
and calculation of divergent integrals : The choice of the scale parameter
which will remain in the divergent objects. In fact, one can use an arbitrary scale,
since the following equations relating them and which can be derived algebraically
are valid.

\b
I_{quad}(m^2_1)=I_{quad}(m^2_2)+(m_1^2-m_2^2)I_{log}(m_2^2)+
\left(\frac{i}{(4\pi )^2}\right)\left[m_1^2-m_2^2-m_1^2ln\left(\frac{m_1^2}
{m^2_2}\right)\right]
\e

\b
I_{log}(m^2_1)=I_{log}(m^2_2)-
\left(\frac{i}{(4\pi )^2}\right)ln\left(\frac{m_1^2}
{m^2_2}\right),
\e

The above relations are valid in Dimensional Regularization an can also be
obtained by a straightforward manipulation of integrals. When used in
connection with physical amplitudes, they correspond to a parametrization
of the freedom we have when separating the finite from the divergent content
of the amplitude. When considering Ward identities
(or renormalization procedures) one resorts to eq.(21) and eq.(22) in order to recover
the adequate mass for that case. Now we treat our divergent integral in such a way to show
explicitly these aspects. First the quadratic divergence. If we maintain the same
mass in the divergent object, $m_t$ for example, we can use the result so obtained,
eq.(15), 
\b
\int_\Lambda \frac{d^4k}{(2\pi )^4}\frac{1}{[(k+k_1)^2 -m_t^2]}=
\left\{I_{quad}(m^2_t)+k_{1\alpha}k_{1\beta}\triangle_{\alpha\beta}(m^2_t)\right\}
\e
But the same integral can written in a completely equivalent way:
\ba
\int_\Lambda \frac{d^4k}{(2\pi )^4}\frac{1}{[(k+k_1)^2-m_t^2]}&=&
\int_\Lambda \frac{d^4k}{(2\pi )^4}\frac{1}
{(k^2-\lambda^2)^2}\nonumber \\
& &-(\lambda^2-m^2_t)\int_\Lambda\frac{d^4k}{(2\pi )^4}\frac{1}
{(k^2-\lambda^2)^2}\nonumber \\
& &+k_{1\alpha}k_{1\beta}\left\{
\int_\Lambda\frac{d^4k}{(2\pi )^4}\frac{4k_\alpha k_\beta}
{(k^2-\lambda^2)^3}-\int_\Lambda\frac{d^4k}{(2\pi )^4}\frac{g_{\alpha\beta}}
{(k^2-\lambda^2)^2}
\right\}
\nonumber \\
& &
\!\!\!\!\!\!\!\!\!\!\!\!\!\!\!\!\!\!\!\!\!\!\!\!\!\!\!\!\!\!\!\!\!\!\!\!\!\!\!\!\!\!\!\!\!\!\!\!\!\!\!\!\!
+\left\{\int\frac{d^4k}{(2\pi )^4}\frac{(k_1^2-m_t^2+\lambda^2)^2}
{(k^2-\lambda^2)^3}
-\int\frac{d^4k}{(2\pi )^4}\frac{(k_1^2+2k_1\cdot k+\lambda^2-m_t^2)^3}
{(k^2-\lambda^2)^3[(k+k_1)^2-m_t^2]}
\right\}
\ea

Again, the right hand side is identical to the left except by the odd factors. 
After obtaining the solution of the finite integral we can write this
\ba
\int_\Lambda \frac{d^4k}{(2\pi )^4}\frac{1}{[(k+k_1)^2-m_t^2]}&=&
\left\{I_{quad}(\lambda^2)+(m_t^2-\lambda^2)I_{log}(\lambda^2)\right.\nonumber \\
& &\left.
+\left(\frac{i}{(4\pi )^2}\right)\left[m_t^2-\lambda^2+m_t^2ln\left(
\frac{\lambda^2}
{m^2_t}\right)\right]\right\}\nonumber \\
& &+k_{1\alpha}k_{1\beta}\triangle_{\alpha\beta}(\lambda^2)
\ea
given the relation (21) the term between brackets is simply $I_{quad}(m_t^2)$ and 
thus we get, for the unambiguous term the same result. Note that with these 
manipulations we can generate another kind of ambiguities (scale ambiguities). 
In this manipulations $\lambda^2$ was taken as an arbitrary mass (scale) but, if we 
want, the value can be chosen equal to the other mass $m_b^2$. The same operations 
can be performed in the logarithmically divergent integral. We can write 
\ba
& &\int_\Lambda \frac{d^4k}{(2\pi )^4}\frac{1}
{[(k+k_1)^2-m_t^2][(k+k_2)^2-m_b^2]}\nonumber \\
& &\;\;\;\;\;\;\;\;\;\;\;\;\;=I_{log}(\lambda^2)-
\left(\frac{i}{(4\pi )^2}\right)\left[
Z_0(m_t^2,m_b^2;(k_1-k_2)^2;\lambda^2)\right]\nonumber \\
& &\;\;\;\;\;\;\;\;\;\;\;\;\;=I_{log}(m^2_b)-
\left(\frac{i}{(4\pi )^2}\right)\left[
Z_0(m_t^2,m_b^2;(k_1-k_2)^2;m_b^2)\right]\nonumber \\
& &\;\;\;\;\;\;\;\;\;\;\;\;\;
=I_{log}(m_t^2)-
\left(\frac{i}{(4\pi )^2}\right)\left[
Z_0(m_t^2,m_b^2;(k_1-k_2)^2;m_t^2)\right]
\ea

The above relations clearly show the role played by the scale parameter, 
left inside the divergent integral. The first equation explicitly shows the
independence of the results on the choice of $\lambda^2$, i.e.,
\b
\frac{\partial I}{\partial \lambda^2}=0.
\e

It is now appropriate to say that the above property will be equally valid 
for unambiguous part of any divergent integral. This is an extra ingredient 
for our forthcoming analysis. At this point differences between our procedure
and BPHZ subtraction scheme become clear. From our point of view any 
identity can be used, which is adequate for the reorganization we have in mind. 
Taylor expansions are one of the possibilities, useful in some restricted cases 
(equal masses, etc.).

After these important remarks we use the results to write $T^{PP}$ as:
\ba
T^{PP}&=&
-2\left\{I_{quad}(m_t^2)+I_{quad}(m_b^2)+[(m_t-m_b^2)^2-(k_1-k_2)^2]
I_{log}(m_t^2)\right.\nonumber \\
& &\left.
-\left(\frac{i}{(4\pi )^2}\right)[(m_t-m_b)^2-(k_1-k_2)^2]\left[
Z_0(m_t^2,m_b^2;(k_1-k_2)^2;m_t^2)\right]\right\}\nonumber \\
& &-2k_{1\alpha}k_{1\beta}\triangle_{\alpha\beta}(m_t^2)
-2k_{2\alpha}k_{2\beta}\triangle_{\alpha\beta}(m_b^2)
\ea

With the same ingredients the case of $T^{SS}$ (eq.(1)), now with different 
masses, can be written in the form

\ba
T^{SS}&=&
2\left\{I_{quad}(m_t^2)+I_{quad}(m_b^2)+[(m_t+m_b^2)^2-(k_1-k_2)^2]
I_{log}(m_t^2)\right.\nonumber \\
& &\left.
-\left(\frac{i}{(4\pi )^2}\right)[(m_t+m_b)^2-(k_1-k_2)^2]\left[
Z_0(m_t^2,m_b^2;(k_1-k_2)^2;m_t^2)\right]\right\}\nonumber \\
& &+2k_{1\alpha}k_{1\beta}\triangle_{\alpha\beta}(m_t^2)
+2k_{2\alpha}k_{2\beta}\triangle_{\alpha\beta}(m_b^2)
\ea

In the last two results we can immediately obtain the corresponding 
equal masses results, eq.(8) and eq.(12). Following the same strategy we obtain for $T^{VS}_\mu$, 
defined by
\b
T^{VS}_\mu =\int \frac{d^4k}{(2\pi )^4}Tr \left\{ \gamma_\mu \frac{1}
{[(k\sla +k\sla_1) -m_t]}\hat{1} \frac{1}{[(k\sla +k\sla _2)
-m_b]}\right\},
\e
with the result

\ba
T^{VS}_\mu &=&
4(k_1-k_2)_\mu\left\{\frac{(m_b-m_t)}{2}I_{log}(m_t^2)\right.\nonumber \\
& &-(m_t+m_b)^2\left(\frac{i}{(4\pi )^2}\right)
\left[Z_1(m_t^2,m_b^2;(k_1-k_2)^2;m_t^2)\right]\nonumber \\
& &\left.
+m_t\left(\frac{i}{(4\pi )^2}\right)\left[
Z_0(m_t^2,m_b^2;(k_1-k_2)^2;m_t^2)\right]\right\}\nonumber \\
& &-2(m_b+m_t)(k_{1}+k_{2})_\nu \triangle_{\nu\mu}(m_t^2).
\ea

Note that the unambiguous terms vanish when the masses are equal, as they should, to 
be compatible with a conservation of corresponding vector current.

Now for $T^{PA}_\nu$,
\b
T^{PA}_\nu =\int \frac{d^4k}{(2\pi )^4}Tr \left\{ \gamma_5 \frac{1}
{[(k\sla +k\sla_1) -m_t]}\gamma_\nu\gamma_5 \frac{1}{[(k\sla +k\sla _2)-m_b]}
\right\}
\e
we have:
\ba
T^{PA}_\nu &=&
-4(k_1-k_2)_\nu\left\{\frac{(m_b+m_t)}{2}I_{log}(m_t^2)\right.\nonumber \\
& &-(m_b-m_t)\left(\frac{i}{(4\pi )^2}\right)
\left[Z_1(m_t^2,m_b^2;(k_1-k_2)^2;m_t^2)\right]\nonumber \\
& &\left.
+m_t\left(\frac{i}{(4\pi )^2}\right)\left[
Z_0(m_t^2,m_b^2;(k_1-k_2)^2;m_t^2)\right]\right\}\nonumber \\
& &+2(m_b-m_t)(k_{1}+k_{2})_\mu \triangle_{\nu\mu}(m_t^2)
\ea

From which the unambiguous result for equal masses becomes apparent.

Now let us consider the two point functions with two Lorentz indices. Firstly
\b
T^{AV}_{\mu\nu} =\int \frac{d^4k}{(2\pi )^4}Tr \left\{ \gamma_\mu \gamma_5 \frac{1}
{[(k\sla +k\sla_1) -m_t]}\gamma_\nu \frac{1}{[(k\sla +k\sla _2)
-m_b]}\right\}
\e

for which we get

\b
T^{AV}_{\mu\nu} =
-4i\varepsilon_{\mu\nu\alpha\beta}\left\{
\frac{k_{1\xi}k_{2\alpha}}{2}\triangle_{\xi\alpha}(m^2_t)+
\frac{k_{2\xi}k_{1\alpha}}{2}\triangle_{\xi\beta}(m^2_t)
\right\}
\e

The above expression vanishes identically if we make the choice ${\cite{ref6}}$
for the internal 
momenta $k_1=\alpha q$, $k_2=(\alpha -1)q$. $T^{AV}_{\mu\nu}$ 
should then contain two Ward identities related to the contractions $(k_1-k_2)_\mu$ 
and $(k_1-k_2)_\nu$ with the amplitude. The only consistent possibility is obtained 
for $T^{AV}_{\mu\nu}=0$.

Now we consider the vector-vector two point function defined by

\b
T^{VV}_{\mu\nu} =\int \frac{d^4k}{(2\pi )^4}Tr \left\{ \gamma_\mu \frac{1}
{[(k\sla +k\sla_1) -m_t]}\gamma_\nu \frac{1}{[(k\sla +k\sla _2)
-m_b]}\right\}.
\e

After Dirac trace we can write the above equation in the convenient form
\b
T^{VV}_{\mu\nu}=T_{\mu\nu}+g_{\mu\nu}[T^{PP}]
\e
where we have introduced the definition:
\b
T_{\mu\nu}=\int \frac{d^4k}{(2\pi )^4} \frac{[(k+k_1)_\mu (k+k_2)_\nu 
+(k+k_1)_\nu (k+k_2 )_\mu]}
{[(k +k_1)^2 -m_t^2][(k +k_2)^2 -m_b^2]}.
\e

The calculation of $T_{\mu\nu}$ is lengthy but straightforward. The result is
(choosing $m_t$ as scale)
\ba
T_{\mu\nu} &=&
\frac{4i}{(4\pi )^2}\left\{
2(q^2g_{\mu\nu}-q_{\mu}q_{\nu})\left[
Z_2(m_t^2,m_b^2;q^2;m_t^2)
-Z_1(m_t^2,m_b^2;q^2;m_t^2)\right]
\right.\nonumber \\
& &\;\;\;\;\;\left.+g_{\mu\nu}(q^2+m_t^2-m_b^2)\left[
Z_0(m_t^2,m_b^2;q^2;m_t^2)
-Z_1(m_t^2,m_b^2;q^2;m_t^2)\right]
\right\}\nonumber \\
& &+4\left\{g_{\mu\nu}[I_{log}(m_t^2)]-\left[\frac{q^2}{6}+\frac{(m_t^2-m_b^2)}
{2}\right]g_{\mu\nu}I_{log}(m_t^2)\right.\nonumber \\
& &\;\;\;\;\;-\left.\frac{q_\mu q_\nu}{3}I_{log}(m_t^2)\right\}+\varphi_{\mu\nu}
\ea
where
\ba
\varphi_{\mu\nu}&=&\frac{k_{2\alpha}k_{2\beta}+k_{1\alpha}k_{1\beta}
+k_{1\alpha}k_{2\beta}}{3}\left\{\Box_{\alpha \beta \mu \nu }(m_t^2)
+g_{\alpha\beta}\triangle_{\mu\nu}(m^2_t)\right.\nonumber \\
& &\;\;\;\;\;\;\;\;\;\;\;\;\;\;\;\;\left.+g_{\alpha\mu}
\triangle_{\beta\nu}(m^2_t)+g_{\alpha\nu}\triangle_{\mu\beta}(m^2_t)\right\}
\nonumber \\
& &-\frac{(k_{1\mu}+k_{2\mu})(k_{1\alpha}+
k_{2\alpha})}{2}\triangle_{\alpha \nu}(m^2_t)+\nonumber \\
& &-\frac{(k_{1\nu}+k_{2\nu})(k_{1\beta}+
k_{2\beta})}{2}\triangle_{\beta \mu}(m^2_t)+\nonumber \\
& &+\nabla_{\mu\nu}(m_t^2)
-\frac{(k_{1}^2+k_{2}^2+m_t^2-m_b^2)}{2}\triangle_{\nu \mu}(m^2_t),
\ea
and we have introduced the definitions
\b
\nabla_{\mu\nu}(m_t^2)=\int_\Lambda \frac{d^4k}{(2\pi )^4}\frac{2k_\mu k_\nu }{(k^2 -m_t^2
)^2}-\int_\Lambda \frac{d^4k}{(2\pi )^4}\frac{g_{\mu\nu}}{(k^2 -m_t^2
)}
\e
and
\ba
\Box_{\alpha \beta \mu \nu }(m_t^2)&=&\int_\Lambda \frac{d^4k}{(2\pi )^4}\frac{
24k_\mu k_\nu k_\alpha k_\beta }{(k^2 -m_t^2 )^4} -g_{\alpha \beta}
\int_\Lambda \frac{d^4k}{(2\pi )^4}\frac{4k_\mu k_\nu }{[(k^2 -m_t^2)^3]}\nonumber \\
& &-g_{\alpha \mu}
\int_\Lambda \frac{d^4k}{(2\pi )^4}\frac{4k_\beta k_\nu }{[(k^2 -m_t^2)^3]}
-g_{\alpha \nu}
\int_\Lambda \frac{d^4k}{(2\pi )^4}\frac{4k_\beta k_\mu }{[(k^2 -m_t^2)^3]}
\ea

Using eqs.(39) and (28) we obtain for $T^{VV}$, after some 
reorganization 

\ba
T_{\mu\nu}^{VV} &=&\Theta_{\mu\nu}
-\frac{4i}{(4\pi )^2}\left\{
\frac{(q^2+m_t^2-m_b^2)}{2q^2}\left[m_t^2-m_b^2+m_b^2ln\left(\frac{m_b^2}
{m_t^2}\right)
\right]\right.\nonumber \\
& &\;\;\;\;\;\;\;\;\;\;\;\;\;\;\;\;\;\;\;\;\;\;\;
\left.-\frac{(m_t-m_b)^2[q^2-(m_t+m_b)^2]}{2q^2}
Z_0(m_t^2,m_b^2;q^2;m_t^2)\right\}g_{\mu\nu}\nonumber \\
& &+4\left\{\frac{[I_{quad}(m_t^2)]-[I_{quad}(m_b^2)]}{2}
-m_t(m_t-m_b)[I_{log}(m_t^2)]\right\}g_{\mu\nu}
+\phi_{\mu\nu}
\ea

where we introduced the definition

\ba
\Theta_{\mu\nu} &=&
\frac{4i}{(4\pi )^2}\left\{
2(q^2g_{\mu\nu}-q_{\mu}q_{\nu})\left[
Z_2(m_t^2,m_b^2;q^2;m_t^2)
-Z_1(m_t^2,m_b^2;q^2;m_t^2)\right]
\right\}\nonumber \\
& &+\frac{4(q^2g_{\mu\nu}-q_{\mu}q_{\nu})}{3}[I_{log}(m_t^2)]
\ea

Which represents the gauge invariant part of the amplitude, and
\b
\phi_{\mu\nu} =\varphi_{\mu\nu} -2k_{1\alpha}k_{1\beta}\triangle_{\alpha\beta}
(m_t^2)-2k_{2\alpha}k_{2\beta}\triangle_{\alpha\beta}
(m_b^2)
\e
which represents the ambiguous part of the amplitude. Now we consider the last 
two point functions we need: $T^{AA}_{\mu\nu}$, defined by:
\b
T^{AA}_{\mu\nu} =\int \frac{d^4k}{(2\pi )^4}Tr \left\{ \gamma_\mu\gamma_5 \frac{1}
{[(k\sla +k\sla_1) -m_t]}\gamma_\nu\gamma_5 \frac{1}{[(k\sla +k\sla _2)
-m_b]}\right\}.
\e

After Dirac trace we write
\b
T^{AA}_{\mu\nu}=T_{\mu\nu}-g_{\mu\nu}[T^{SS}]
\e

Then, using eqs.(39) and (29) we obtain
\ba
T_{\mu\nu}^{AA} &=&\Theta_{\mu\nu}
+\frac{4i}{(4\pi )^2}\left\{
\frac{[q^2-(m_t-m_b)^2](m_b+m_t)^2)}{2q^2}Z_0(m_t^2,m_b^2;q^2;m_t^2)\right\}g_{\mu\nu}
\nonumber \\
& &-\frac{4[q^2-(m_t-m_b)^2](m_b+m_t)^2)}{2q^2}[I_{log}(m_t^2)]g_{\mu\nu}
\nonumber \\
& &+\frac{4(m_t^2-m_b^2)}{2q^2}\left[I_{quad}(m_b^2)-I_{quad}(m_t^2)
\right]g_{\mu\nu}
+\phi_{\mu\nu}
\ea

The other possible two point functions, not considered explicitly are all
identically zero ($T^{SP}$, $T^{VP}_\mu$, $T^{AS}_\mu$). 

Finally, for use in our future considerations, let us consider the one 
point functions, defined as
\b
T^{V}_\mu (m)=\int \frac{d^4k}{(2\pi )^4}Tr \left\{ \gamma_\mu \frac{1}
{[(k\sla +l\sla ) -m]}\right\}
\e

or, given by
\ba
T^{V}_\mu (m)&=&4\left\{(-)l_\alpha \left(\nabla_{\alpha\mu}(m^2)\right)
-\frac{l_\alpha l_\beta l_\nu}{3}\left[\Box_{\alpha \beta \mu \nu }(m^2)
\right]\right.\nonumber \\
& &\left.-\frac{l_\alpha l_\beta l_\nu}{3}\left[\nabla_{\beta \nu }(m^2)
\right]+\frac{l^2 l_\nu}{3}\left[\triangle_{\nu \mu }(m^2)
\right]+l_\mu l_\alpha l_\nu \left[\triangle_{\alpha \nu }(m^2)\right]
\right\}
\ea

The other one point functions vanish identically ($T^P(m)=T^A_\mu (m)=0$).

Before ending this section we call attention for the fact that the ambiguous 
character of {\it all} considered amplitudes are coefficients of one of the three 
above defined objects $\triangle_{\alpha\beta}$, $\nabla_{\alpha\beta}$ and 
$\Box_{\alpha\beta\mu\nu}$.

\section{Ambiguities and Ward Identities}

In the results presented in the previous section we have only made use of 
identities {\it at the level of the integrand} and integration without 
restriction in the finite parts. The last statement can be put in other 
words: the effect of the regularization on this finite integrals is neglected, 
as done in ref.[4].

Notice that in all results, ambiguous terms appear as coefficients of only three 
relations between divergent integrals of the same degree of divergence. It is 
a simple matter to obtain the results of ref.[4] from ours. For example if we use 
some regularization (like sharp cutoff) in the relation $\triangle_{\alpha\beta}
(m_t^2)$.

\b
\triangle_{\alpha\beta}^{Reg}(m_t^2)=\int \frac{d^4k}{(2\pi )^4}
\left\{\frac{4k_\alpha k_\beta}{(k^2-m_t^2)^3}-\frac{g_{\alpha\beta}}
{(k^2-m_t^2)^3}
\right\}G_\Lambda (k^2,\Lambda^2)
\e
we can immediately use $k_\mu k_\nu =\frac{k^2g_{\mu\nu}}{4}$ and get, taking 
$\Lambda^2>>m^2$;
\b
\triangle_{\alpha\beta}^{Reg}(m_t^2)=\left(\frac{i}{(4\pi )^2}\right)
\left[\frac{-1}{2}\right].
\e

And the results for $T^{SS}$ and $T^{PP}$ of ref.[4] are obtained. The same 
procedure can be applied to the other two relations, 
$\Box_{\alpha \beta \mu \nu }$ and $\nabla_{\mu\nu}$, with corresponding 
results. The full contact with the results of ref.[4] is made by expressing 
$I_{log}$ and $I_{quad}$ in their regularization scheme. As discussed before, this
procedure obviously leads to ambiguities and violations of gauge invariance.
We can ask ourselves, at this point, what could be done with our 
expression to avoid such problems. We remind the reader that we have not made 
use of any regularization prescription so far. In order to decide this 
we invoke the Ward identities which {\it must be} satisfied in order to preserve
gauge invariance. They are exact relations between the various two point 
amplitudes and can be directly established from their definition through simple 
algebraic manipulations in the trace operation. Let us consider one example. We 
contract $T^{VS}_\mu$ with $(k_1-k_2)_\mu =q_\mu$
\b
q^\mu T^{VS}_{\mu} =\int \frac{d^4k}{(2\pi )^4}Tr \left\{ \hat{1} \frac{1}
{[(k\sla +k\sla_1) -m_b]}(k\sla_1-k\sla_2) \frac{1}{[(k\sla +k\sla _2)
-m_t]}\right\}
\e

Now we use the identity
\b
(k\sla_1-k\sla_2)=(k\sla_1+k\sla -m_t)-(k\sla_2+k\sla -m_b)+(m_t-m_b)
\e
in order to get
\b
q^\mu T_\mu^{VS}=(m_t-m_b)T^{SS}+T^{S}(m_b)-T^S(m_t)
\e
where we have identified the two point functions 
$T^{SS}$, eq.(1), and scalar one point functions, eq.(13). By means of this 
procedure one can easily get other relations:
\b
q^\mu T_{\mu\nu}^{VV}=(m_t-m_b)T^{SV}_\nu +T^V_\nu (m_b)-T^V_\nu (m_t)
\e
\b
q^\mu T_{\mu\nu}^{AA}=(m_t+m_b)T^{PA}_\nu +T^V_\nu (m_b)-T^V_\nu (m_t)
\e
\b
q^\mu T_{\mu}^{AP}=-(m_t+m_b)T^{PP}+T^S(m_b)-T^S(m_t)
\e
\b
q^\mu T_{\mu\nu}^{AV}=(m_t+m_b)T^{PV}_\nu +T^A_\nu (m_b)-T^A_\nu (m_t)
\e
\b
q^\mu q^\nu T_{\mu\nu}^{VV}=(m_t-m_b)^2T^{SS}+(m_t-m_b)[T^S(m_b)-T^S(m_t)]
\e
\b
q^\mu q^\nu T_{\mu\nu}^{AA}=-\left[(m_t+m_b)^2T^{PP}+(m_t+m_b)[-T^S(m_b)+T^S(m_t)]
\right]
\e

These relationships are exact and must be satisfied by any regularization 
scheme employed in the calculations. Otherwise the method is inconsistent. We 
then proceed to investigate general conditions to be satisfied by any consistent 
regularization prescription. As will become clear in what follows it is quite 
simple to find the set of conditions to be imposed from the point of view of 
symmetries and ambiguities on a regularization scheme.

The above conclusions can be drawn already from the analysis of the 
calculated divergence, namely, the one point vector function. As a consequence of Furry's 
theorem $T^V_\mu (m^2)$ must vanish identically. A regularization prescription 
not capable of fulfilling the identity will immediately violate gauge
invariance in QED's vacuum polarization tensor (eq.(56) for equal masses). 
We could take $l=0$ in eq.(50) and this would be individually satisfied for 
$T^V_\mu (m^2)$. However in the Ward identities we have the difference between two 
of these amplitudes with different momenta which cannot be simultaneously 
put to zero. Thus, in order to have $T^V_\mu (m^2)=0$, an adequate regularization 
prescription should satisfy
\b
\left\{
\begin{array}{ll}
\Box_{\alpha \beta \mu \nu }^{Reg}(m^2)=0 \\
\triangle_{\alpha\beta}^{Reg}(m^2)=0 \\
\nabla_{\alpha\beta}^{Reg}(m^2)=0
\end{array}
\right.
\e

The above relations can therefore be viewed as minimal consistency conditions 
for regularizations prescriptions. They are necessary and sufficient in order 
to simultaneously inforce symmetry preservation in the
perturbative analysis. As discussed before the physical root of the above 
consistency conditions is translational invariance of the original lagrangean.
Let us then assume the conditions eq.(62) as part of our strategy. All 
ambiguities are thus avoided. It is therefore crucial to prove that at least
one such regularization scheme exists. This has been done in ${\cite{ref12}}$,
and shown here in Appendix A for the sake of completeness.

The relevant question now is: are the conditions eq.(62) sufficient to preserve 
the Ward identities? The answer to this question is non trivial, as can be 
gathered from an analysis of the identity related to $T^{VS}_\mu$ 
eq.(55), as compared to the result obtained from the direct calculation of 
such amplitude eq.(31). In order to satisfy the Ward identity, we need the difference 
between two quadratically divergent integrals, since the two $T^S$ one point 
functions are part of the identity. Note, however, that in the result 
obtained for $T^{VS}_\mu$ quadratic divergences are completely absent. However, 
when we use the (exact) mathematical relation between the functions $Z_1$ and 
$Z_0$,

\ba
Z_1(m_t^2,m_b^2,q^2;m_b^2)&=&
Z_0(m_t^2,m_b^2;q^2;m_t^2)\frac{(q^2+m_t^2-m_b^2)}{2q^2}\nonumber \\
& &+\frac{1}{2q^2}\left[m_t^2-m_b^2+m_b^2ln\left(\frac{m_b^2}{m_t^2}\right)
\right]
\ea
and the (also exact) scale relations, eq.(21) and eq.(22), we get
\b
T^{VS}_\mu = \frac{(k_1-k_2)_\mu}{(k_1-k_2)^2}\left\{
(m_t-m_b)[T^{SS}]+T^S(m_b)-T^S(m_t)\right\}
\e
by direct identification of $T^{SS}$, $T^{S}(m_b)$ and $T^{S}(m_t)$ 
given in eqs.(29) and eq.(15), respectively. The same procedure allows us to 
write $T^{PA}_\nu$ in 
the following form
\b
T^{PA}_\nu = \frac{(k_1-k_2)_\nu}{(k_1-k_2)^2}\left\{
-(m_t+m_b)[T^{PP}]+T^S(m_b)-T^S(m_t)\right\}
\e
Also $T^{VV}_{\mu\nu}$ can be cast in the form 
\b
T^{VV}_{\mu\nu} = \Theta_{\mu\nu}+(m_t-m_b)\frac{g_{\mu\nu}}{q^2}\left\{
(m_t-m_b)[T^{SS}]+T^S(m_b)-T^S(m_t)\right\}
\e
and
\b
T^{AA}_{\mu\nu} = \Theta_{\mu\nu}+(m_t+m_b)\frac{g_{\mu\nu}}{q^2}\left\{
-(m_t+m_b)[T^{PP}]+T^S(m_b)-T^S(m_t)\right\}
\e

From eqs.(65), (66) and (67) the corresponding Ward identities follow
immedialtely. It is important to stress that all these relations are exact
and are a direct consequence of the validity of the relations
eq.(21) and eq.(22) and {\it consistency conditions} eq.(62).

Note that the adoption of the consistency conditions implies in the validity of the 
relations
\b
\int_\Lambda \frac{d^4k}{(2\pi )^4}\frac{2k_\mu k_\nu }{[(k^2 -m^2)^2]}=
g_{\mu\nu}[I_{quad}(m^2)]
\e
\b
\int_\Lambda \frac{d^4k}{(2\pi )^4}\frac{4k_\mu k_\nu }{[(k^2 -m^2)^3]}=
g_{\mu\nu}[I_{log}(m^2)]
\e
\b
\int_\Lambda \frac{d^4k}{(2\pi )^4}\frac{24k_\mu k_\nu k_\alpha k_\beta}
{[(k^2 -m^2)^3]}=
[g_{\mu\nu}g_{\alpha\beta}+g_{\mu\alpha}g_{\nu\beta}+g_{\mu\beta}
g_{\nu\alpha}]I_{log}(m^2)
\e

These relations in particular show that all the divergent content of the one loop 
amplitudes of the NJL model are in one of the forms $I_{quad}$ or $I_{log}$. 
Therefore the only divergent objects which need to be calculated are those 
two. This fact reduces the role played by an eventual (consistent) 
regularization method to that of furnishing a parametrization of such integrals.

Of course calculations beyond one loop bring in other structures like
overlapping divergences etc. It is a simple matter to extend the method in
order to deal with them. In the present work, as mentioned before, we
restricted ourselves to one loop calculations.

It is important to remember that such parametrization must be consistent with
the scale relations and therefore should satisfy
\b
\left\{
\begin{array}{ll}
\frac{\partial I_{quad}(\lambda^2)}{\partial \lambda^2}=I_{log}(\lambda^2)\\
\frac{\partial I_{log}(\lambda^2)}{\partial \lambda^2}=\left(\frac{i}
{(4\pi )^2}\right)\left(\frac{-1}{\lambda^2}\right)
\end{array}
\right. .
\e

One possible such parametrization could be
\b
\left\{
\begin{array}{ll}
I_{quad}(\lambda^2)=\left(\frac{i}
{(4\pi )^2}\right)[-\Lambda^2+\lambda^2+\lambda^2ln\left(\frac{\Lambda^2}
{ \lambda^2}\right)+\beta_0\lambda^2+\delta_0]\\
I_{log}(\lambda^2)=\left(\frac{i}
{(4\pi )^2}\right)[ln\left(\frac{\lambda^2}
{ \Lambda^2}\right)+\beta_0]
\end{array}
\right.
\e
where $\Lambda^2$ play a role of a regularization parameter (cut off) and $\beta_0$ 
and $\delta_0$ are constants.

Although the above relations are useful and valid in general for any method, in the 
specific context of the NJL model they are not necessary, since the gap 
equation relates the constituent quark mass $m$ directly to $I_{quad}(m^2)$ 
and the decay constant $g_2$ directly to $I_{log}(m^2)$. This procedure yields 
a direct relation between divergent quantities and phenomenological physical 
quantities.

It is nowadays a current point of view that in order to have a complete 
definition of a nonrenormalizable model the specification of a regularization 
scheme is necessary ${\cite{ref16}}$. The results extracted from our analysis allow for
the conclusion that the NJL within the prescription used exhibits its full
predictive power. It was shown to be consistent, free of ambiguities and
symmetry violations. The gauge invariance of the W-boson propagator is only a
consequence oh this consistency, but it is not the only one.

In order to conclude this section a comment about the solution proposed by 
ref.${\cite{ref6}}$ is in order: In ref.${\cite{ref6}}$ the proposed solution is via dispersion
relations. In this case the starting point is the imaginary part of the amplitudes
as dictated by Cutkosky's rules and the real part is constructed through
dispersion relations. In our results the imaginary part is contained in the
$Z_k$ functions, given for example by ${\cite{ref16}}$:
\b
Im \left\{Z_0(m_t^2,m_b^2,q^2;m_t^2)\right\}=
2\pi \Theta \left(q^2-(m_t+m_b)^2\right)\sqrt{q^2 -(m_t+m_b)^2}
\frac{\sqrt{q^2 -(m_t-m_b)^2}}{2q^2}
\e

an immediate check reveals the compatibility with Cutkosky's rules in such a 
way that if we had used those rules to construct the amplitudes we would get the 
same results for the cut off independent part of the results of ref.${\cite{ref6}}$.

\section{Conclusions}

In the present work we revisited the questions raised on ref.${\cite{ref6}}$ with respect
to consistent regularization schemes for treating the gauged NJL model, from a 
different point of view. In particular we can obtain their results as a particular 
case.

From our analysis it is possible to conclude that in order to maintain the NJL full
predictive power it is crucial that the eventually used regularization prescription
satisfies:
\b
\left\{
\begin{array}{ll}
\Box_{\alpha \beta \mu \nu }^{Reg}(m^2)=0 \\
\triangle_{\alpha\beta}^{Reg}(m^2)=0 \\
\nabla_{\alpha\beta}^{Reg}(m^2)=0
\end{array}
\right.
\e

The above relations, as discussed, are enough to eliminate  all possible 
ambiguities and symmetry violations. In this case the only remaining 
divergent objects are of the form $I_{quad}(\lambda^2)$ and $I_{log}(\lambda^2)$. 
Moreover, in order to explicitly verify the Ward identities it is also necessary
that the algebraic relations eqs.(21) and (22) are preserved. These relations
have been derived algebraically and in the present context their importance
is due the fact that the explicit verification of several Ward identities
would not be possible were these relations not valid. They have been called
a manifestation of scale invariance of the calculated amplitude since
$\lambda^2$ plays the role of a {\it scale}. This interpretation becomes solid
if one studies the Renormalization Group (RG) of Quantum Electrodynamics
at the one loop level using the present prescription. As we know, at the
one loop level the only divergence which appears is $I_{log}(m^2)$, if we
choose $m^2$ as the renormalization point. The freedom we have in choosing a
different renormalization point is equivalent to the use of eq. (22). It is a
simple matter to check that the R.G. coefficients are obtained in a regularization
independent fashion once eq.(22) is used in the appropriate amplitudes in QED
to parametrize the freedom one has in choosing the renormalization point ${\cite{ref17}}$.

In this case it is natural to assume that the conditions eq.(74) are related to
some fundamental property of Quantum Field Theories. In fact, it can be shown 
that they are a consequence of translational invariance ${\cite{ref12}}$.
Therefore, 
a consistent method should automatically incorporate such conditions as in 
the case of Dimensional Regularization.

For the specific case of the NJL model one important consequence of the present 
strategy is that all its divergent content at the one loop level can be expressed 
in terms of two basic objects $I_{log}$ and $I_{quad}$, which are usually related 
to physical quantities; the constituent quark mass (gap equation) and the 
coupling constant $g_2$. The predictions of the model depend only on parameters 
of the model itself.

The present prescription for the manipulation and calculation of divergent 
amplitudes ${\cite{ref15}}$ is not restricted to the NJL model and can be also
applied to effective theories in general and in renormalization programs. It 
should be emphasized that wherever Dimensional Regularization applies we get
essentially the same results, provided the divergent objects are written 
according to the method (an investigation of the existence of 4
dimensional regularizations which obey this rule has also been proved
${\cite{ref12}}$).

One other important ingredient of the present prescription is the systematization of 
the finite content of all amplitudes in terms of the $Z_k$ functions ${\cite{ref15}}$.
These functions have a much wider applicability. Indeed, the finite part of 
{\it any } one loop amplitude can be cast into this form. They 
enormously simplify the analysis of the Ward identities, relevant physical 
limits and directly emphasize aspects related to unitarity.

\section{Acknowledgment}

We are indebted to A. L. Mota for fruitful discussions and for the
demonstration in Appendix A, part of his Ph.D. thesis.

\section{Appendix A}

As discussed in ref. ${\cite{ref12}}$ and briefly summarized here,
our consistency conditions at one loop level are but a consequence
of translational invariance. Let us consider the free Green's function
of the theory and require that it is translationally invariant

\begin{equation}
S(x,x^{\prime })=\int \frac{d^4k}{(2\pi )^4}\frac{e^{ik(x-x^{\prime })}}{%
\gamma .k-m}
\end{equation}

The  ''translated ''Green's function is given by

\begin{equation}
S_{\alpha q}(x,x^{\prime })=\int \frac{d^4k}{(2\pi )^4}\frac{e^{i(k+\alpha
q)(x-x^{\prime })}}{\gamma .(k+\alpha q)-m}  \label{Salfaq}
\end{equation}

When acting on a test function, there should be no difference between $S$ and $%
S_{\alpha q}$, so we require that $S=S_{\alpha q}$, for arbitrary $\alpha q$.

\begin{equation}
\int S(x,x^{\prime })J(x^{\prime })d^4x^{\prime }=\int S_{\alpha
q}(x,x^{\prime })J(x^{\prime })d^4x^{\prime }
\end{equation}

The generating functional of the free theory, defined in the standard way :
\begin{equation}
Z_o[J]=N\exp \{-i\int J(x)S(x,x^{\prime })J(x^{\prime
})\,\,\,\,d^4x\,\,\,\,d^4x^{\prime }\}  \label{ZoJ}
\end{equation}
is independent of the parameter $\alpha $. For the generating functional
of the interacting theory we have
\begin{equation}
Z[J]=\exp \{-i\int L_{int}(-i\frac \delta {\delta J(z)})d^4z\}Z_o[J]
\label{ZJ}
\end{equation}

All the amplitudes evaluated in the present work can be obtained by means of the
convenient functional derivatives of the above generating functional.
It is straightfoward to show that when we use the translated Green's
function $S_{\alpha q}$ instead of $S$, we obtain the amplitudes
with the arbitrary momentum routing.

Now we argue that once $Zo[J]$ is $\alpha $ independent, so must $Z[J]$ be.
However since we are dealing with ill defined quantities, a regularization
scheme must be defined :
\begin{eqnarray}
\int S_{\alpha q}(x,x^{\prime })J(x^{\prime })d^4x^{\prime } &=&\int
d^4x^{\prime }\int \frac{d^4k}{(2\pi )^4}\frac{e^{i(k+\alpha q)(x-x^{\prime
})}}{\gamma .(k+\alpha q)-m}\times  \\
\times \int \frac{d^4p}{(2\pi )^4}e^{ipx\prime }\stackrel{\symbol{126}}{J}%
(p) &=&\int \frac{d^4k}{(2\pi )^4}\frac{e^{i(k+\alpha q)x}}{\gamma
.(k+\alpha q)-m}\stackrel{\symbol{126}}{J}(k+\alpha q)=  \nonumber \\
&=&\int \frac{d^4k}{(2\pi )^4}\exp \{\alpha q^\mu \frac \partial {\partial
k^\mu }\}(\frac{e^{ikx}}{\gamma .k-m}\stackrel{\symbol{126}}{J}(k)) 
\nonumber
\end{eqnarray}
where we have introduced the shift operator $\exp \{\alpha q^\mu \frac
\partial {\partial k^\mu }\}$. Expanding the shift operator we get that the
integrals proportional to $\alpha ^n$ are surface terms, and for $%
J(x^{\prime })$ being a adequate test function, they will vanish.
But on the improper integrals $S_{\alpha q}$ will act over a distribution,
\begin{equation}
\int d^4x\int d^4x^{\prime }\,\,\,S_{\alpha q}(x,x^{\prime })D(x,x^{\prime })
\label{ImpInt}
\end{equation}
where $D(x,x^{\prime })$ is a distribution, typically the delta function or
products of single particle Green's functions :
\begin{equation}
\int d^4x\int d^4x^{\prime }\,\,\,S_{\alpha q}(x,x^{\prime })D(x,x^{\prime
})=\int \frac{d^4k}{(2\pi )^4}\exp \{\alpha q^\mu \frac \partial {\partial
k^\mu }\}(\frac 1{\gamma .k-m}\stackrel{\symbol{126}}{D}(k))  \label{SalfaqD}
\end{equation}

For $D(x-x^{\prime })=\delta ^4(x-x^{\prime })$, we have $\stackrel{\symbol{%
126}}{D}(k)=1$, and thus :

\begin{eqnarray}
S_{\alpha q}(0) &=&\int \frac{d^4k}{(2\pi )^4}\exp \{\alpha q^\mu \frac
\partial {\partial k^\mu }\}(\frac 1{\gamma .k-m})=  \label{Salfaq0} \\
&=&\int \frac{d^4k}{(2\pi )^4}\frac 1{\gamma .k-m}+\alpha q^\mu \int \frac{%
d^4k}{(2\pi )^4}\frac \partial {\partial k^\mu }\{\frac 1{\gamma .k-m}\}+ 
\nonumber \\
&&+\alpha ^2q^\mu q^\nu \int \frac{d^4k}{(2\pi )^4}\frac{\partial ^2}{%
\partial k^\mu \partial k^\nu }\{\frac 1{\gamma .k-m}\}+...  \nonumber
\end{eqnarray}

The above equation will be $\alpha $ independent if and only if relations
(74) are satisfied. One possible regularization scheme which implements
this feature is obtained by replacing the particle's Green's function
by the sequence of functions which define the distribution, i. e.
\begin{equation}
\int S(x,x^{\prime })J(x^{\prime })d^4x^{\prime }=\mathrel{\mathop{\lim
}\limits_{n\longrightarrow \infty }}\int S_n(x,x^{\prime })J(x^{\prime
})d^4x^{\prime }  \label{Snx}
\end{equation}
where
\begin{equation}
S_n(x,x^{\prime })=\int \frac{d^4k}{(2\pi )^4}\frac{e^{ik(x-x\prime )}}{%
\gamma .k-m}\exp [-\frac{\sigma k^2}{4n^2}]
\end{equation}
where $\sigma $ is a parameter with the appropriated dimension and signal to
make the sequence $\{S_n(x,x^{\prime })\}$ to converge to the distribution $%
S(x,x^{\prime })$.

\end{document}